\documentclass{article}

\usepackage[nonatbib,preprint]{neurips_2020}

\usepackage[utf8]{inputenc} 
\usepackage[T1]{fontenc}    
\usepackage{hyperref}       
\usepackage{url}            
\usepackage{booktabs}       
\usepackage{wrapfig}
\usepackage{amsfonts}       
\usepackage{amsmath,amsthm,amssymb}
\usepackage{scalerel}
\usepackage{mathrsfs}
\usepackage{enumitem}
\usepackage{algorithm}
\usepackage[noend]{algpseudocode}
\usepackage{mcr}
\usepackage{graphicx}
\usepackage{subcaption}
\usepackage{epstopdf}
\usepackage{todonotes}
\usepackage{multirow}
\usepackage[normalem]{ulem}
\useunder{\uline}{\ul}{}
\usepackage{wrapfig}

\title{Multi-Label Learning to Rank through Multi-Objective Optimization}

\author{Debabrata Mahapatra\thanks{Amazon intern}\\
	National University of Singapore\\
	\texttt{debabrata@u.nus.edu} 
	\And
	Chaosheng Dong\thanks{Corresponding author} \\
	Amazon.com Inc\\
	\texttt{chaosd@amazon.com} 
	\And
	Yetian Chen \\
	Amazon.com Inc\\
	\texttt{yetichen@amazon.com} 
	\And
	Deqiang Meng\\
	Amazon.com Inc\\
	\texttt{deqiangm@amazon.com} 
	\And
	Michinari Momma\textsuperscript{$\dagger$}\\
	Amazon.com Inc\\
	\texttt{michi@amazon.com} 
}

\def \bx {\boldsymbol{x}}

\def \bb {\boldsymbol{b}}

\def \balpha {\boldsymbol{\alpha}}

\def \bG {\boldsymbol{G}}

\def \bl {\boldsymbol{l}}

\begin{document}
	\maketitle
	
	\begin{abstract}
		Learning to Rank (LTR) technique is ubiquitous in the Information Retrieval system nowadays, especially in the Search Ranking application. The query-item relevance labels typically used to train the ranking model are often noisy measurements of human behavior, e.g., product rating for product search. The coarse measurements make the ground truth ranking non-unique with respect to a single relevance criterion. To resolve ambiguity, it is desirable to train a model using many relevance criteria, giving rise to Multi-Label LTR (MLLTR). Moreover, it formulates multiple goals that may be conflicting yet important to optimize for simultaneously, e.g., in product search, a ranking model can be trained based on product quality and  purchase likelihood to increase revenue. 
		In this research, we leverage the Multi-Objective Optimization (MOO) aspect of the MLLTR problem and employ recently developed MOO algorithms to solve it. Specifically, we propose a general framework where the information from labels can be combined in a variety of ways to meaningfully characterize the trade-off among the goals. Our framework allows for any gradient based MOO algorithm to be used for solving the MLLTR problem. We test the proposed framework on two publicly available LTR datasets and one e-commerce dataset to show its efficacy. 
	\end{abstract}

	\section{Introduction}
	Research in Learning to Rank (LTR) has exploded in the last decade. It can be attributed to the increasing availability of labeled data for query-item relevance, either through manually labeling or tracking user behavior.
	In LTR, a scoring function is trained to score the retrieved items for ranking. 
	Originally, LTR was developed to use only one relevance criterion for training. However, owing to the limitations of such uni-dimensional approach, e.g., subjectivity and noise in relevance articulation, inability to incorporate multiple goals, a multi-dimensional approach for relevance is adopted in Multi-Label Learning to Rank (MLLTR) \cite{10.1145/1963405.1963459}. 
	The multidimensional aspect of MLLTR poses a fundamental challenge: different relevance criteria can be conflicting. For example, in web search, two conflicting criteria could be considering the user-history and  increasing serendipitous items in the top results. Due to this conflict, it is virtually infeasible to find a scoring function that simultaneously optimizes for all relevance criteria, thus requires a trade-off among them. 
	
	\begin{wrapfigure}{R}{0.35\textwidth}
		\centering
		\includegraphics[width=0.99\linewidth, page=5]{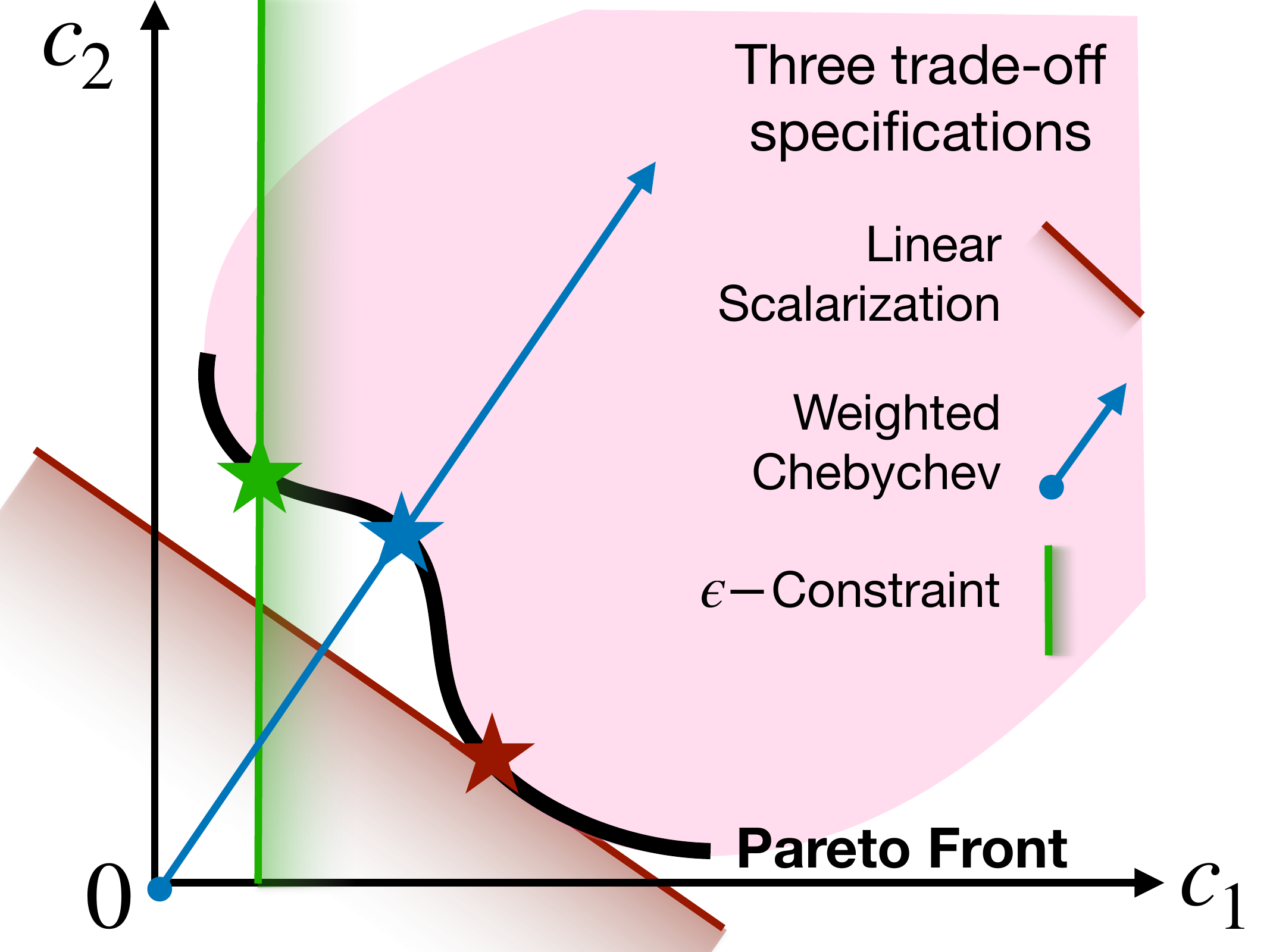}
		\caption{Three  trade-off specifications investigated for MLLTR.}
		\label{fig:3_to}
	\end{wrapfigure}
	
	The field of Multi-Objective Optimization (MOO) naturally models the trade-offs among different objectives through the \textit{Pareto Frontier} (PF), which is a set of (possibly infinitely many) non-dominated solutions (i.e., ranking functions in our context). The rich history of MOO research has introduced several ways for specifying a trade-off \cite{DBLP:books/daglib/0021267}, e.g., setting priorities on objectives (Linear Scalarization, Weighted Chebyshev, etc.) or constraining them ($\epsilon-$Constraint method), that could lead to a unique non-dominated solution.  
	However, recent studies on MLLTR have mainly focused on approximating the PF, whose individual solutions may not associate to a particular trade-off \cite{10.1145/3366423.3380122}, or at best to one type of trade-off \cite{momma2019multi}. 
	Although approximating the entire PF without associating a solution with a trade-off specification is appealing, it is of little use for MLLTR because the final consumable presented to a user is one ranked list of items, not an ensemble of rankings. 
	In contrast, we bring together disciplines of MOO and MLLTR to not only approximate the PF, but also to find the scoring functions associated with different types of trade-off specifications that would be more applicable in real world deployment. 

	\subsection{Our Contributions} 
	Our contributions are four-fold: 1) We pose the MLLTR problem as an MOO problem and develop a general framework under which any first order gradient based MOO algorithm can be incorporated. 2) We investigate three types of trade-off specifications as shown in Figure \ref{fig:3_to}. We analyze their pros and cons  and one can decide when to use which method according to the application. 3) We reveal common challenges when directly applying these MOO methods, and propose a smoothing technique to resolve them.  4) We test the proposed MLLTR framework on two publicly available datasets and one e-commerce dataset, to compare MOO approaches of obtaining trade-off solutions. The revised algorithms significantly improve the performance on these datasets, indicating that we have a realistic way to build MLLTR models that may benefit many production systems in industry.

	\subsection{Related Work} 
	Several research have incorporated multiple relevance labels in information retrieval systems, e.g., in web search and recommendation \cite{10.1145/1963405.1963459, 10.1145/2009916.2009933, 10.1145/2187836.2187894, 10.1145/2124295.2124350,10.1145/2348283.2348385}, and product search \cite{10.1145/2396761.2398671,10.1145/3077136.3080838}. These traditional methods can be classified into two categories: model aggregation, where individually trained models are combined to give the final ranking, and label aggregation, where the relevance labels are combined to give a single ranking model. The state of the art Stochastic Label Aggregation (SLA) method \cite{10.1145/3366423.3380122} shows equivalence to Linear Scalarization (LS). For recommendation application, \cite{10.1145/3298689.3346998} proposed a framework for MOO based MLLTR that, although guarantees to find non-dominated solutions, does not account for trade-offs. For product search application,  \cite{momma2019multi} proposed many relevance criteria, and developed an $\epsilon-$Constraint MOO algorithm that allows for trade-off specification as upper bounds of all objectives but one.  
	
	Recently, many gradient based MOO algorithms have been developed for the application of Multi-Task Learning (MTL) \cite{sener2018multi,lin2019pareto} that approximate the PF.  \cite{pmlr-v119-mahapatra20a} developed an EPO algorithm for MTL that guarantees to find the solutions corresponding to trade-off specifications defined by the objective priorities. \cite{michipmtl2022} developed a WC-MGDA algorithm for MTL that achieves the same guarantees and can improve over arbitrary reference models. \cite{gong2021automatic} proposed  DBGD algorithm, an $\epsilon-$Constraint type of method that allows for trade-off specification as upper bounds of all objectives but one.  In our MLLTR framework, we facilitate trade-off specification through many of the MOO methods mentioned above, starting from the classic LS to the modern ones like EPO. 
	
	Another related line of research in ranking also considers a multi-objective/multi-task learning framework. However, it uses only one relevance label and adds auxiliary objectives/tasks to force the ranking function yields results satisfying specific criteria, such as scale calibration \cite{51402}, fairness \cite{singh2019policy,morik2020controlling}, and diversity \cite{li2020cascading}. Moreover, to the best of our knowledge, it essentially uses the classic LS approach to solve the yielded multi-objective optimization problem. In contrast, our paper emphasizes the exploration of applying advanced MOO algorithms for  MLLTR.

	\section{Background}\label{sec:background}
	\subsection{Learning to Rank}
	\setlength{\abovedisplayskip}{1pt}
	\setlength{\belowdisplayskip}{1pt}
	Let $\mathbb{Q}$ be the set of all possible queries and $\mathbb{D}$ be the set of all documents or items. For a given query $q\in\mathbb{Q}$, let ${D^q = \{d_i\}_{i=1}^{n_q} \subset \mathbb{D}}$ be the subset of $n_q$ matched items. Let a query-item pair $(q, d_i)$ be represented by a $p$-dimensional feature vector $\mathbf{x}^q_i \in \mathbb{R}^p$. The goal of LTR is to learn a parametric scoring function $f_{\boldsymbol{\theta}}:\mathbb{R}^p \rightarrow \mathbb{R}$ that can assign a score $s^q_i$ to each $(q, d_i)$ pair from its corresponding vector representation, i.e., $\mathbf{x}^q_i \mapsto s^q_i$. The items can then be ranked in descending order of scores.
	
	For a $(q, d_i)$ pair, we denote the relevance label as $y^q_i \in \mathbb{Y}$,
	The training dataset for LTR consists of several queries: $ 	\mathcal{D}_\mathrm{LTR} = \Big\{\big\{ \left(\mathbf{x}^q_i, y^q_i\right)\big\}_{i=1}^{n_q} \Big\}_{q=1}^m $,
	where $m$ is the number of queries and $n_q$ is the number of data points in each query group. 
	
	For a query $q$, let the output of a scoring function $f_{\boldsymbol{\theta}}$ for all the matched items in $D^q$ be represented by a score vector $\mathbf{s}^q\in\mathbb{R}^{n_q}$. Similarly, let the corresponding relevance labels be denoted by the vector $\mathbf{y}^q \in \mathbb{R}^{n_q}$. The training cost is given by
	\begin{align}\label{eq:cost_ltr}
	c(\boldsymbol{\theta}) = \frac{1}{m} \sum_{q=1}^m \ell(\mathbf{s}^q, \mathbf{y}^q), \quad \text{where } s^q_i= f_{\boldsymbol{\theta}} (\mathbf{x}^q_i)
	\end{align}
	for all $i \in [n_q]\! =\! \{1, 2, \cdots, n_q\}$, and the per-query loss $\ell(\mathbf{s}^q, \mathbf{y}^q)$ quantifies the extent to which the ordering of scores disagrees with that of the relevance labels.
	
	In the pair-wise approach of LambdaMART cost \cite{Burges2010FromRT}, the event that one item $d_i$ is more relevant than another $d_j$ w.r.t. $q$, denoted by $d_i \rhd_{\!q} d_j$, is probabilistically modeled as $ \mathrm{P}(d_i \rhd_{\!q} d_j)  =  \frac{1}{1 + e^{- \sigma (s^q_i - s^q_j)}} $,
	where $\sigma$ controls the spread of  the Sigmoid function. The per-query loss $\ell$ in \eqref{eq:cost_ltr} is constructed from the log-likelihood ($\ell\ell$) of $\boldsymbol{\theta}$ given the (presumably independent) observations in the training data:
	\begin{align}\label{eq:ranknet_loss}
	\!\!\ell(\mathbf{s}^q\!, \mathbf{y}^q)= -\ell\ell(\boldsymbol{\theta}|\mathcal{D}^q)  = \!\!\sum_{(i,j)\in I^q} |\Delta NDCG(i,j)|  \cdot \log\left(1+e^{-\sigma (s^q_i - s^q_j)}\right),
	\end{align}
	where $\mathcal{D}^q\!=\!\{(\mathbf{x}^q_i, y^q_i)\}_{i=1}^{n_q}$ is data pertaining to the matched items $D^q$, $I^q = \left\{(i,j) \in [n_q]^2 \; \middle|\; y^q_i > y^q_j\right\}$ consists of item pairs having a strict relevance order, and $ \Delta NDCG(i,j)  $ is the change of the NDCG value when two items $ i $ and $ j $ swap their rank positions \cite{Burges2010FromRT}.
	
	The scoring function is modeled by GBM \cite{friedman2001greedy} with $N$ decision trees: $ f^N_{\boldsymbol{\theta}}(\mathbf{x}) = T_{\theta^0}(\mathbf{x}) -\sum_{t=1}^{N-1} \eta_t T_{\theta^t}(\mathbf{x}) $,
	where $\eta_t$ is the learning rate, $T_{\theta^t}$ is the $t^\text{th}$ tree, and the full model parameter is $\boldsymbol{\theta} = \{\theta^t\}_{t=0}^{N-1}$. On the $t^\text{th}$ iteration,
	the tree $T_{\theta^t}$ is learnt from the following training data:
	\begin{align}\label{eq:data_gbm}
	\mathcal{D}_{T_{\theta^t}} =
	\left\{\left\{\left(\mathbf{x}_i^q,  { \partial c} / {\partial s_i^q}\right)\right\}_{i=1}^{n_q}\right\}_{q=1}^m,
	\end{align}
	where the labels are gradients of cost w.r.t. the scores. 
	In other words, instead of updating $f_{\boldsymbol{\theta}}$ in the parameter space, it is updated in the function space of trees: $f^{t+1}_{\boldsymbol{\theta}} = f^t_{\boldsymbol{\theta}} - \eta_t T_{\theta^t}$.
	
	The function space update of GBM suffices to treat the cost as a function of scores rather than the  parameters $\boldsymbol{\theta}$. Henceforth, we consider the cost $c:\mathbb{R}^M\rightarrow \mathbb{R}$ as a function of $\mathbf{s}$ , and  rewrite \eqref{eq:cost_ltr} as 
	\begin{align}\label{eq:cost_ltr_score}
	c(\mathbf{s}) = \frac{1}{m} \sum_{q=1}^{m} \ell(\mathbf{s}^q, \mathbf{y}^q).
	\end{align}

	\subsection{Learning to Rank from multiple relevance labels}
	In MLLTR, different relevance criteria are measured, providing multiple labels for each query-item pair. The goal of MLLTR is still the same as that of LTR: to learn a scoring function $f_{\boldsymbol{\theta}}$ that assigns a scalar value to each $(q, d_i)$ pair. 
	
	The labels for $(q, d_i)$ are ${y^q_{ik} \in \mathbb{Y}_k}$ for ${k = 1, \cdots, K}$, where $K$ is the number of relevance criteria. Similar to LTR, each label set $\mathbb{Y}_k$ could be either discrete or continuous, endowed with a total ordering relation. The training dataset for MLLTR is denoted by
	\begin{align}\label{eq:data_mlltr}
	\mathcal{D}_\mathrm{MLLTR} = \Big\{\big\{ \left(\mathbf{x}^q_i, y^q_{i1}, \cdots, y^q_{iK} \right)\big\}_{i=1}^{n_q} \Big\}_{q=1}^m.
	\end{align}
	
	Each relevance criterion has a training cost. Therefore, in MLLTR, the cost is a vector valued function: $ \mathbf{c}(\mathbf{s}) = \left[c_1(\mathbf{s}), \cdots, c_K(\mathbf{s})\right]^T $,
	naturally making it an MOO problem. 
	
	\subsection{Multi-Objective optimization}\label{sec:moo}
	In MOO, the cost function ${\mathbf{c}:\mathbb{R}^{M} \rightarrow
		\mathbb{R}^K}$ is a mapping from the \textit{solution space} $\mathbb{R}^{M}$ to the \textit{objective space}~$\mathbb{R}^K$. 
	
	We use the cone of positive orthant, i.e., ${\mathbb{R}^K_+ := \left\lbrace \mathbf{c}\in \mathbb{R}^K \;\middle|\; c_k \ge 0, \; \forall\ k \in [K]\right\rbrace}$, to define a partial ordering relation.
	For any two points ${\mathbf{c}^1, \mathbf{c}^{2} \in \mathbb{R}^K}$, we write ${\mathbf{c}^1 \succcurlyeq \mathbf{c}^2}$, if $\mathbf{c}^1$ lies in the positive cone pivoted at $\mathbf{c}^2$, i.e., $\mathbf{c}^1 \in \left\{ \mathbf{c}^2 + \mathbf{c} \;\middle|\; \mathbf{c} \in \mathbb{R}^K_+ \right\}$.
	In other words, $\mathbf{c}^1 \succcurlyeq \mathbf{c}^2 \iff \mathbf{c}^1-\mathbf{c}^2 \in \mathbb{R}^K_+$, making $c^1_k \ge c^2_k$, $\forall k\in [K]$. We define $\mathbf{c}^1 \succ \mathbf{c}^2$  when there is at least one $k$ for which $c^1_k > c^2_k$, i.e., $\mathbf{c}^1 \neq \mathbf{c}^2$.

	
	For minimization, a solution $\mathbf{s} \in \mathbb{R}^{M}$ is said to be non-dominated or \textit{Pareto optimal}, if there exists no other solution $\mathbf{s}' \in \mathbb{R}^{M}$ such that $\mathbf{c}(\mathbf{s}) \succ \mathbf{c}(\mathbf{s}')$. 
	We call the set of all non-dominated solutions the Pareto optimal set. 
	The image of this Pareto set under the function $\mathbf{c}$ is the \textit{Pareto Frontier} (PF).

	\begin{figure*}[t]
		\centering
		\begin{subfigure}{0.24\textwidth}
			\centering
			\includegraphics[width=\linewidth, 
			page=1]{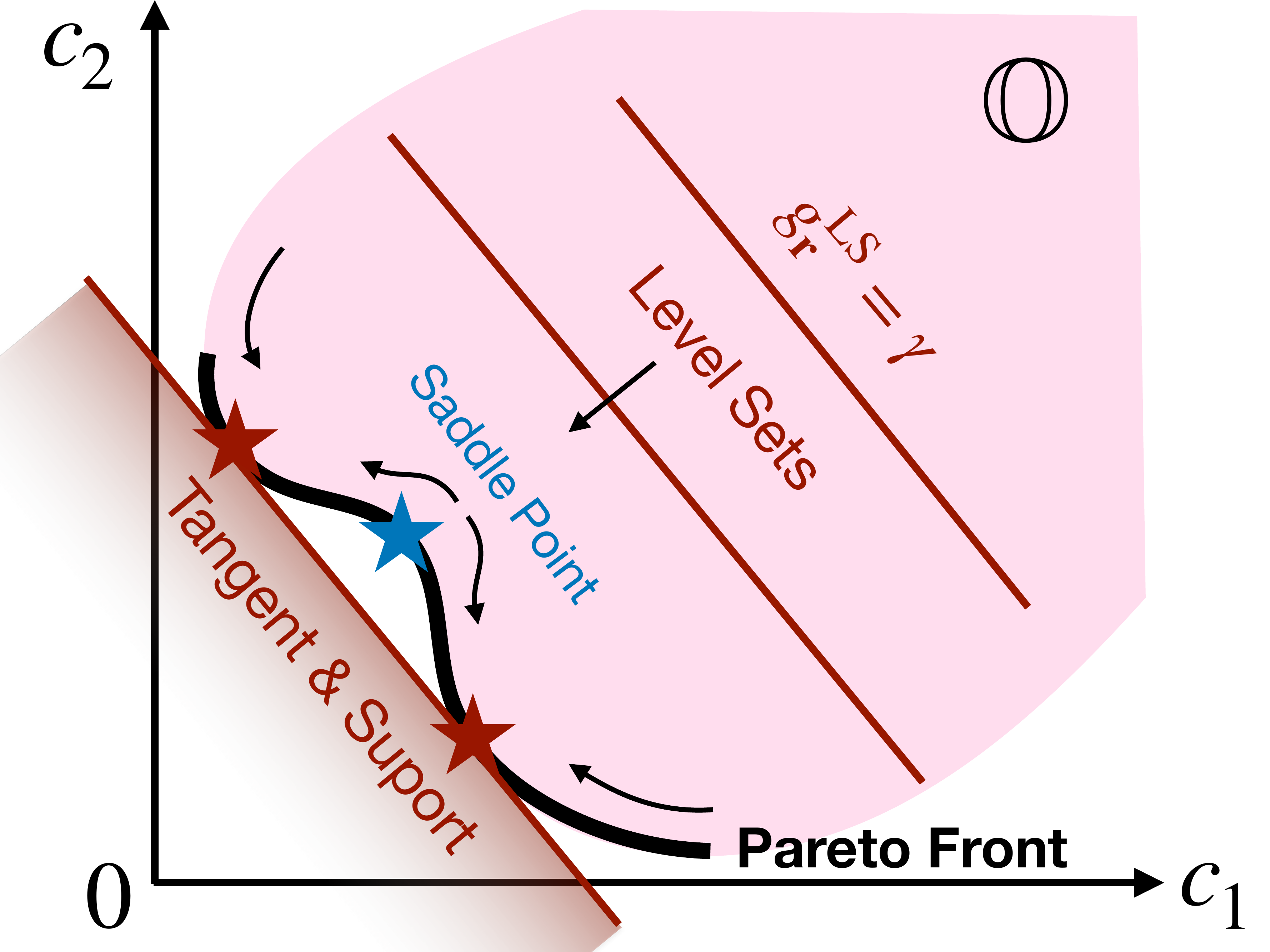}
			\caption{\label{fig:linscal} LS \& SLA}
		\end{subfigure}
		\begin{subfigure}{0.24\textwidth}
			\centering
			\includegraphics[width=\linewidth,
			page=2]{figure/illustrations}
			\caption{\centering\label{fig:chbscal} WC  \& EPO}
		\end{subfigure}	
		\begin{subfigure}{0.24\textwidth}
			\centering
			\includegraphics[width=\linewidth, 
			page=3]{figure/illustrations}
			\caption{\label{fig:wc-mgda} WC-MGDA}
		\end{subfigure}
		\begin{subfigure}{0.24\textwidth}
			\centering
			\includegraphics[width=\linewidth, 
			page=4]{figure/illustrations}
			\caption{\label{fig:econst} $\epsilon-$Constraint methods}
		\end{subfigure}
		\caption{\label{fig:tradeoffs} Illustration of trade-off specifications. 
			\eqref{fig:linscal} shows how LS can have non-unique Pareto optimal points.
			\eqref{fig:chbscal} shows WC can attain the blue optimum by minimizing $g_\mathbf{r}^\mathrm{WC}$ and  illustrates how EPO works. \eqref{fig:wc-mgda} shows WC-MGDA can find Pareto optima better than the arbitrary reference point $ \mathbf{b} $. \eqref{fig:econst} shows  that  $\epsilon-$Constraint can find different Pareto optima by constraining the cost $C_1$.}
	\end{figure*}
	
	\section{A framework for Multi-Label Learning to Rank}\label{sec:framework}
	

	\textbf{Multi-Gradient Combination}
	The MLLTR cost function gives rise to $K$ score-gradients, $\nabla_{\!\mathbf{s}}c_k$ for $k\in[K]$. However, for training the GBM based scoring function, the $t^\text{th}$ decision tree requires exactly one score-gradient as labels in its training data \eqref{eq:data_gbm}, not $K$ score-gradients. Although the cost is upgraded to become a vector valued function in MLLTR, the scoring function remains a scalar valued function. We combine the $K$ score-gradients  as
	\begin{align}
	\boldsymbol{\lambda} = \sum_{k=1}^{K} \alpha_k \nabla_{\!\mathbf{s}}c_k, \quad \text{s.t.} \ \sum_{k=1}^{K} \alpha_k = 1, \quad \boldsymbol{\alpha} \in \mathbb{R}^K_+,
	\end{align}
	where  $\boldsymbol{\lambda} \in \mathbb{R}^M$ are the labels for training the trees in GBM and $\boldsymbol{\alpha}$ are combination coefficients.

	\subsection{Linear Scalarization based methods}
	\textbf{Linear Scalarization (LS):}
	The MOO cost is converted to a scalar cost $ g_\mathbf{r}^\textrm{LS}(\mathbf{s}) = \sum_{k=1}^K r_kc_k(\mathbf{s})$,
	where $\mathbf{r}\in \mathbb{R}^K_+$ represents preferences/priorities given to the costs. 
	
	\textit{Gradient Combination:} 	It remains static throughout the iterations
	\begin{align}\label{eq:ls_alpha}
	\boldsymbol{\alpha}=\mathbf{r} / \|\mathbf{r}\|_1.
	\end{align}
	
	Although LS is simple, specifying trade-offs by elements in the dual space has limitations. If any of the costs is a non-convex function, i.e., the range $\mathbb{O}$ becomes a non-convex set, LS can not guarantee to reach all points in the PF by varying the preferences \cite{boyd_vandenberghe_2004},  as illustrated in Figure\ref{fig:linscal}.
	

	\textbf{Stochastic Label Aggregation (SLA):}
	One gradient is randomly chosen following the distribution:
	\begin{align}\label{eq:sla}
	\alpha_{k} = \begin{cases}
	1, \quad \text{if}\ k = \overline{K}, \\
	0, \quad \text{otherwise},
	\end{cases}
	\text{for  } k\in[K]
	\end{align}
	where $\overline{K}$ is a categorical random variable over the K indices with $\mathbf{r}/\|\mathbf{r}\|_1$ as its probability distribution.
	In other words, the $\overline{K}^\text{th}$ label is used for training. The expected cost of SLA is the same as that of LS \cite{10.1145/3366423.3380122}. Thus, \textbf{SLA can be seen as a special type of LS.}
	
	\subsection{Preference direction based methods}
	\subsubsection{Weighted Chebyshev  (WC)}
	In WC, the vector valued cost is scalarized to 
	\begin{align}\label{eq:cs}
	g^\textrm{WC}_\mathbf{r} (\mathbf{s}) = \max_{k\in[K]} r_k c_k(\mathbf{s}).
	\end{align}
	In general, the solution $\mathbf{s}_\mathbf{r}^* = \min_\mathbf{s} g_\mathbf{r}^\textrm{WC}(\mathbf{s})$ satisfies $ r_1 c_1(\mathbf{s}^*_\mathbf{r}) = r_2 c_2(\mathbf{s}^*_\mathbf{r}) = \cdots = r_K c_K(\mathbf{s}^*_\mathbf{r}) $ \cite{DBLP:books/daglib/0021267}.
	which can be deduced by analyzing the level sets, illustrated in Figure\ref{fig:chbscal}. This 
	makes the trade-off specification between the objectives stricter than the penalty approach in the LS. 
	
	\textit{Gradient Combination:}
	Only the gradient of maximum relative objective value is chosen:
	\begin{align}\label{eq:cs_alpha}
	\alpha_k = 
	\begin{cases}
	1, \ \text{if } \ k = k^*,  \\
	0, \ \text{otherwise},
	\end{cases}
	\text{s.t. }  k^*=\arg\max_{k\in[K]} r_kc_k(\mathbf{s}).
	\end{align}
	
	The objective vector value is proportional to the $\mathbf{r}^{-1}$ ray as illustrated in Figure\ref{fig:chbscal}. 
	This trade-off specification guarantees that Pareto optimal points in the PF can be reached by varying the preferences, even when the objectives are non-convex. 
	However, in practice, the strict trade-off requirement hinders the progress in cost value reduction. When optimizing with a step size (i.e., learning rate), the iterate $\mathbf{c}^t$ (cost at $t^\text{th}$ iteration) oscillates around $\mathbf{r}^{-1}$ ray.
	

	\subsubsection{Exact Pareto Optimal Search (EPO)}
	In EPO \cite{pmlr-v119-mahapatra20a, mahapatra2021exact}, the trade-off specification is the same as that of WC. Therefore, most properties of WC are inherited. However, to overcome the limitations of WC, its gradient combination is designed to avoid oscillations around the $\mathbf{r}^{-1}$ ray. 
	
	\textit{Gradient Combination:} 
	The coefficients are obtained by solving a quadratic program:
	\begin{align}\label{eq:epo_alpha}
	\min_{\boldsymbol{\alpha}\in \mathbb{R}^K_+} \ \|\mathrm{C}^T\mathrm{C}\boldsymbol{\alpha} - \mathbf{a}\|_2^2, \ \
	\text{s.t.}  \sum_{k=1}^{K} \alpha_k = 1, 
	\end{align}
	where $\mathrm{C}\in \mathbb{R}^{M\times K}$ is the matrix with $K$ gradients in its column, and $\mathbf{a}$ is an \textit{anchor direction} in the objective space that determines the first order change in cost vector:  ${\mathbf{c}^{t+1} - \mathbf{c}^t \approx  \delta\mathbf{c} = \mathrm{C}^T\mathrm{C}\boldsymbol{\alpha}}$ from Taylor series expansion of $\mathbf{c}(\mathbf{s}^t - \mathrm{C}\boldsymbol{\alpha})$.
	Here, $\mathbf{a}$ is determined by
	\begin{align}\label{eq:epo2_alpha}
	\mathbf{a} = \begin{cases}
	\mathbf{c}^t  - \frac{\langle \mathbf{c}^t, \mathbf{r}^{-1}\rangle}{\|\mathbf{r}^{-1}\|_2}\mathbf{r}^{-1}, \quad \text{if} \; \mathbf{c}^t \;\text{is far}, \\
	\mathbf{r}^{-1}, \quad \text{otherwise}.
	\end{cases}
	\end{align}
	When $\mathbf{c}^t$ is far (w.r.t. cosine distance) from $\mathbf{r}^{-1}$ ray, the anchor is orthogonal to the $\mathbf{r}^{-1}$ ray and directs towards it, as illustrated in Figure\ref{fig:chbscal}.
	On the other hand, when $\mathbf{c}^t$ is near $\mathbf{r}^{-1}$ ray, we move the cost along the $\mathbf{r}^{-1}$ ray avoiding oscillations. 
	\subsubsection{Weighted Chebyshev MGDA (WC-MGDA)}
	In WC-MGDA algorithm  \cite{michipmtl2022}, the trade-off specification is similar  to that of WC method, but the SOCP formulations are designed to avoid the shortcomings of WC, i.e., through the preferences over the objectives. WC-MGDA  aims to build models that are closer or better than the reference model. 
	
	\textit{Gradient Combination:} 
	The coefficients are obtained by:
	\begin{equation}\label{eq:wcm-dual}
	\max_{\boldsymbol{\alpha}\in \mathbb{R}^K_+, \bx  \in R^{n}, \gamma}  ~ \balpha^\intercal (\mathbf{r} \odot (\bl(\bx) - \bb)) - u \gamma ~~ \text{s.t.} ~ \sum_{k=1}^{K} \alpha_k = 1, , ~ \|
	\bG_\mathbf{r}  \balpha \|_2 \le \gamma,
	\end{equation}
	where $\bb$ is the loss of the reference model, and $\bG_\mathbf{r} \equiv \text{diag} (\sqrt{\mathbf{r}}) \bG \text{diag} (\sqrt{\mathbf{r}}) $. Here, $ \bG = (\sqrt{\mathrm{C}^T \mathrm{C})}$. 
	
	WC-MGDA jointly solves WC and MGDA to ensure achieving both preference alignment and Pareto Optimality. While the WC problem tries to find solutions by minimizing weighted $\ell_\infty$, the norm minimization ensures Pareto Optimality. 
	\subsubsection{Evaluation metric for preference direction based MLLTR}
	\begin{wrapfigure}{R}{0.25\textwidth}
		\centering
		\includegraphics[width=1\linewidth]{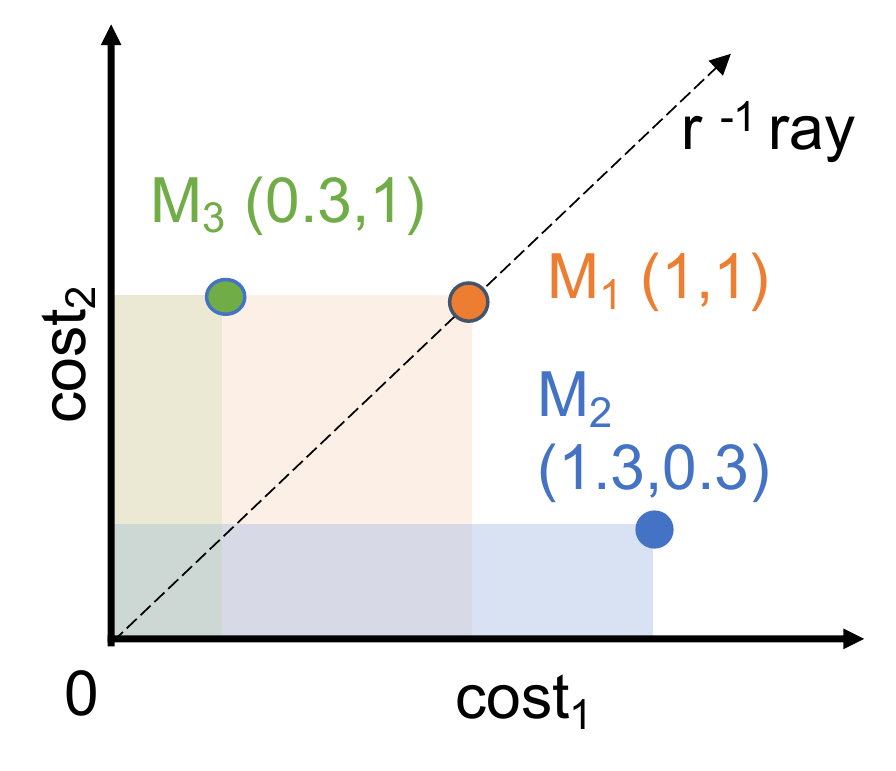}
		\caption{MWL example}
		\label{fig:eval-moo-pd}
	\end{wrapfigure}
	To quantify the performance on preferencde based MLLTR, we use the objective function of WC \eqref{eq:cs}ne, which exactly captures alignment with the $\mathbf{r}^{-1}$-ray and is referred to as maximum weighted loss (MWL).
	Fig. \ref{fig:eval-moo-pd} illustrates a prototypical case
	with 3 models. In terms of MWL, $M_1$ and $M_3$ are the same,
	although $M_3$ dominates $M_1$, and better than $M_2$. Between
	$M_1$ and $M_3$, we use the volume of intersection between the
	negative orthant (VNO) pivoted by each model and
	$\mathbb{R}^K_{+}$ (color shaded area in
	Fig. \ref{fig:eval-moo-pd}) as a tiebreaker. Note, VNO should
	always be used as a tie breaker when the difference in MWL is
	insignificant. Thus, $M_2$ is better than $M_1$ due to VNO.

	\subsection{$\epsilon-$Constraint (EC) methods}
	\subsubsection{$\epsilon-$Constraint Augmented Lagrangian (EC-AL)}
	In this method, the MOO problem is transformed into 
	\begin{align}\label{eq:ec}
	\min_{\mathbf{s} \in \mathbf{R}^M} \ c_{k_{p}}(\mathbf{s}) \quad
	\text{s.t.} \  c_{k}(\mathbf{s}) \leq \epsilon_k, \ \text{for} \ k \in [K] - \{k_{p}\},
	\end{align}
	where one cost $k_p$ is treated as the primary cost and the rest $K-1$ costs are restricted to satisfy an upper bounded constraint given by the $\epsilon_k$. 
	
	\textit{Gradient Combination:}
	\cite{momma2019multi} proposed an augmented Lagrangian form of \eqref{eq:ec} as
	\begin{align}\label{eq:ec_lambdamart}
	\max_{\boldsymbol{\alpha}}\min_{\mathbf{s}}\mathcal{L}(\mathbf{s}, \boldsymbol{\alpha}) = c_{k_p} (\mathbf{s})  + \sum_{k\in [K]_{p}}\alpha_k (c_k(\mathbf{s}) - \epsilon_k),
	\end{align}
	where $[K]_p = [K] - \{k_p\}$.
	At iteration $t$, $  \boldsymbol{\alpha} $ is decided according to a proximal update strategy
	\begin{align}\label{eq:ec_alpha}
	\alpha_k^t = \begin{cases}
	\mu (c^t_k - \epsilon_k) + \alpha^{t-1}_k, \quad \text{if}\ c^t_k - \epsilon_k \geq 0, \\
	0, \quad \text{otherwise}.
	\end{cases}
	\end{align}
	for $ k \in [K] - \{k_p\}$, where $\mu>0$ is a large value, and $\alpha^{t-1}_k$ is the coefficient of the previous iteration. Coefficient of a secondary objective is non-zero only when its constraint is violated.
	\subsubsection{$\epsilon-$Constraint  Dynamic Barrier Gradient Descent (EC-DBGD)}
	The trade-off specification is the same as that of EC-AL, i.e., through the upper bounds on secondary objectives.
	The coefficients are obtained by  solving the following convex quadratic program \cite{gong2021automatic}:
	\begin{equation}\label{eq:dbgd}
	\min _{\boldsymbol{\alpha}, \alpha_{k_p}=1} ~~
	\frac{1}{2}  \left\|
	\mathrm{C}\boldsymbol{\alpha}
	\right\|_2^{2}
	-\sum_{k\in [K]_{p}} \alpha_k
	\phi_{k}\left(\mathbf{s}\right)
	,
	\end{equation}
	where $\phi_{k}$ is a control function associated with constraint $c_{k_p} (\mathbf{s})$ for $ k \in  [K]_{p}$.
	
	
	
	We illustrate three types of trade-off specifications in Figure \ref{fig:tradeoffs}, and summarize the training process in algorithm \ref{alg:mg_combination}, the MOO methods in algorithm \ref{alg:get_coefficients}.
	\begin{figure}[t]
		\begin{minipage}{0.55\linewidth}
			\begin{algorithm}[H]
				{\scriptsize \caption{{\scriptsize MLLTR by GBDT}}
					\label{alg:mg_combination}
					\textbf{Input}: $\mathcal{D}_\mathrm{MLLTR}$ from \eqref{eq:data_mlltr}, $N$, learning rate(s)  $\eta$\\
					\textbf{Parameter}: GBM configurations, $combinator$
					\begin{algorithmic}[1] 
						\State Set $f_{\boldsymbol{\theta}}^0 = T_{\theta^0}$ (Usually, this is set to $0$)
						\For{$t \gets 1$ to $N-1$}
						\State Get $\mathbf{s}$: $s_i^q = f_{\boldsymbol{\theta}}^t(\mathbf{x}_i^q)$  \ for all ($q, d_i$) pair, costs $\mathbf{c}^t$ and gradients $\mathrm{C}^t$
						\State $\boldsymbol{\alpha}$ = \Call{GetCoefficients}{$\mathbf{c}^t,\mathrm{C}^t$, $combinator$}
						\State Prepare data $\mathcal{D}_{T_{\theta}^t}$ as \eqref{eq:data_gbm} but with labels $\boldsymbol{\lambda} = \mathrm{C}^t\boldsymbol{\alpha}$. Fit $\mathbf{T}_{\theta^t}$ to $\mathcal{D}_{T_{\theta}^t}$
						
						\State Update Scoring function: $f^t_{\boldsymbol{\theta}} = f^{t-1}_{\boldsymbol{\theta}} - \eta_t \mathbf{T}_{\theta^t}$
						\If{$\|\mathbf{c}^{t} - \mathbf{c}^{t-1}\|\approx 0$} Break
						\EndIf
						\EndFor
						\State \textbf{return} $f^t_{\boldsymbol{\theta}}$
				\end{algorithmic}}
			\end{algorithm}
		\end{minipage}
		\begin{minipage}{0.45\linewidth}
			\begin{algorithm}[H]
				{\scriptsize \caption{{\scriptsize Multi-Gradient Coefficient from MOO}}
					\label{alg:get_coefficients}
					\begin{algorithmic}[1] 
						\Function{GetCoefficients}{$\mathbf{c}$, $\mathrm{C}$, $combinator$}
						\If {$combinator$ = LS} $\boldsymbol{\alpha}$ from \eqref{eq:ls_alpha}
						\ElsIf{$combinator$ = SLA} $\boldsymbol{\alpha}$ from  \eqref{eq:sla}
						\ElsIf{$combinator$ = WC} $\boldsymbol{\alpha}$ from  \eqref{eq:cs_alpha}
						\ElsIf{$combinator$ = EPO} $\boldsymbol{\alpha}$ from  \eqref{eq:epo_alpha}
						\ElsIf{$combinator$ = WC-MGDA} $\boldsymbol{\alpha}$ from  \eqref{eq:wcm-dual}
						\ElsIf{$combinator$ = EC-AL} $\boldsymbol{\alpha}$ from \eqref{eq:ec_alpha} 
						\ElsIf{$combinator$ = EC-DBGD} $\boldsymbol{\alpha}$ from \eqref{eq:dbgd} 
						\EndIf
						\State \textbf{return} $\boldsymbol{\alpha}$
						\EndFunction
				\end{algorithmic}}
			\end{algorithm}
		\end{minipage}
	\end{figure}

	\section{Experiments}\label{sec:experiments}
	\subsection{Datasets and experimental settings}
	\textbf{{Microsoft Learning to Rank dataset:}}
	We test MOO methods on Microsoft Learning to Rank web search
	dataset (MSLR-WEB30K) \cite{DBLP:journals/corr/QinL13}. Each
	query-url pair is represented by a $136$ dimensional feature
	vector, and 5-level relevance judgment (Rel) is given as the
	original level.  To construct multiple labels, we followed
	\cite{DBLP:journals/corr/abs-2002-05753}, and used four
	\footnote{\textit{Query-URL Click Count} (Click), \textit{URL
			Dwell Time} (Dwell), \textit{Quality Score} (QS) and
		\textit{Quality Score2} (QS2)} of its $136$ features
	as additional relevance labels that are removed when training
	to avoid target leak. 
	We selected all 10 pairs of labels for bi-ojective cases. For
	tri-objective cases, we choose 6 triplets.
	For preference based methods, we generate equi-angular rays in
	the cost space between single objective baselines, and for EC
	models, we generate equi-distance upperbounds between 0 and
	single objective baseline. 5 and 25 $\mathbf{r}^{-1}$ rays are
	generated for bi-objective and tri-objective cases,
	respectively. 
	We tuned the hyperparameters of GBM model by optimizing Rel only, and selected $600$ trees and $0.25$ learning rate,
	by the grid search
	on validation NDCG@5. We repeated the experiment for Folds 1-3,
	and got metrics for analysis.

	\textbf{e-commerce dataset:}
	We test MOO methods on one e-commerce search dataset collected in 2021. This dataset is similar to that used in \cite{momma2019multi}.  Each query is associated with a set of products impressed by customers in a search session,  anusd query-product dependent features such as product sales, textual matches, etc.), as well as customer’s purchase decision. We sampled \textasciitilde 10K queries for training and evaluation, and created the following five labels: 1) a binary target of a product being purchased or not; 2) historical purchases of a products in past 3 months; 3) relevance quality score between queries and products \cite{sementicsim}; 4) brand appealing score of product (probability that a given brand would be judged "high quality" by a human);  5) delivery speed of a product (binary label of whether a product can be delivered in 2 days). From them, we created 8 pairs for bi-objective case and 6 triplets for tri-objective case. We use 300 trees and set the learning rate to be $0.47$ that is selected by random search. To generate preference and constraints, we followed the same strategy mentioned in MSLR-WEB30K. We repeated 3 randomizations for collecting data for analysis.
	
	\textbf{{Yahoo dataset:}} We also run experiment on Yahoo dataset. See Appendix \ref{appendix:tab:yahoo} for the results.
	
	\begin{figure*}[htb]
		\centering
		\begin{subfigure}{0.49\textwidth}
			\includegraphics[page=1, width=\linewidth]{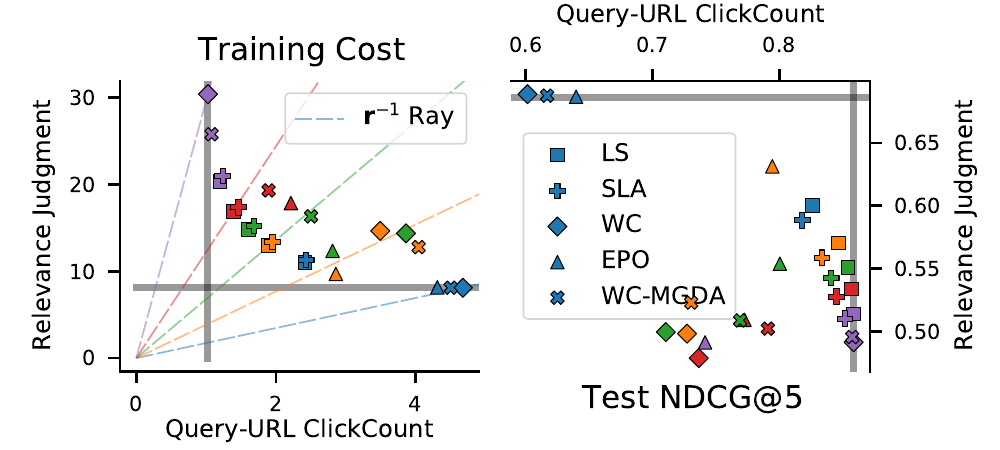}
			\caption{\label{fig:be_qs-qs2} LS/SLA and
				preference based MOO methods. }
		\end{subfigure}
		\begin{subfigure}{0.49\textwidth}
			\includegraphics[page=1, width=\linewidth]{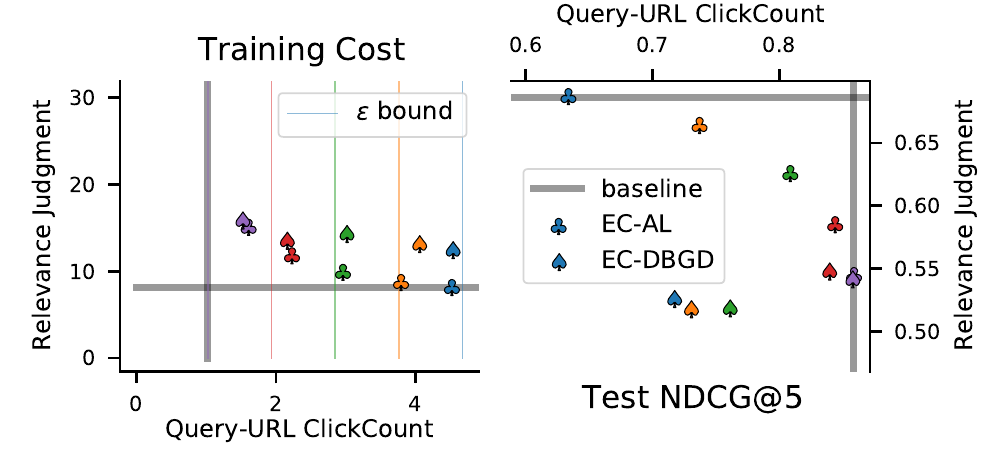}
			\caption{\label{fig:be_qs-rj} $\epsilon$-Constraint methods. }
		\end{subfigure}
		\caption{\label{fig:initial}
			Initial results of bi-objective experiments on
			MSLR-WEB30K \cite{DBLP:journals/corr/QinL13}
			dataset. 
			Colored lines and points represent different
			trade-off specifications and the corresponding
			solutions, respectively.
		}
	\end{figure*}
	
	\subsection{Initial experiment on MSLR dataset}\label{sec:iniexp}
	As an initial experiment to apply existing methods in their
	original form, we ran all methods including the linear
	weighting methods (LS, SLA), preference based methods (WC,
	EPO, WC-MGDA), and EC methods (EC-AL, EC-DBGD) on (Click, Rel) pair on the MLSR dataset. Figure
	\ref{fig:initial} shows the result.
	Surprisingly, while the simpler baselines, LS and EC-AL, performs well and
	seems to achieves PO, other methods are
	inferior to them. For SLA, it is dominated by LS for most of
	the cases in both cost and NDCG results (i.e., square and plus
	points of the same colors in Figure
	\ref{fig:initial}). Further, performance of other methods are quite unstable, and inferior to LS and EC-LA.
	
	To understand this, we plot cost curve for several models
	in Figure \ref{fig:cost_curves}. LS is the only
	method that has smooth behavior in the figure.
	SLA is a stochastic version of LA and
	non-smooth changes are
	visible, which causes inferior
	performance in PO.
	For WC, the oscillation is expected, as it chooses one label that
	have maximum weighted cost. However, EPO, which is
	designed to avoid oscillations, does show similar behavior.
	Clearly,
	the mechanism of EPO is broken
	and even small disruption would cause
	oscillations and hence performance degradations. Same issues
	exist for all pairs (and all datasets in this paper), and
	also observed for WC-MGDA and EC-DBGD.

	\subsection{Remedy by moving average (MA)}
\begin{figure*}[htb]
		\centering
		\begin{subfigure}{0.25\textwidth}
			\includegraphics[width=\linewidth]{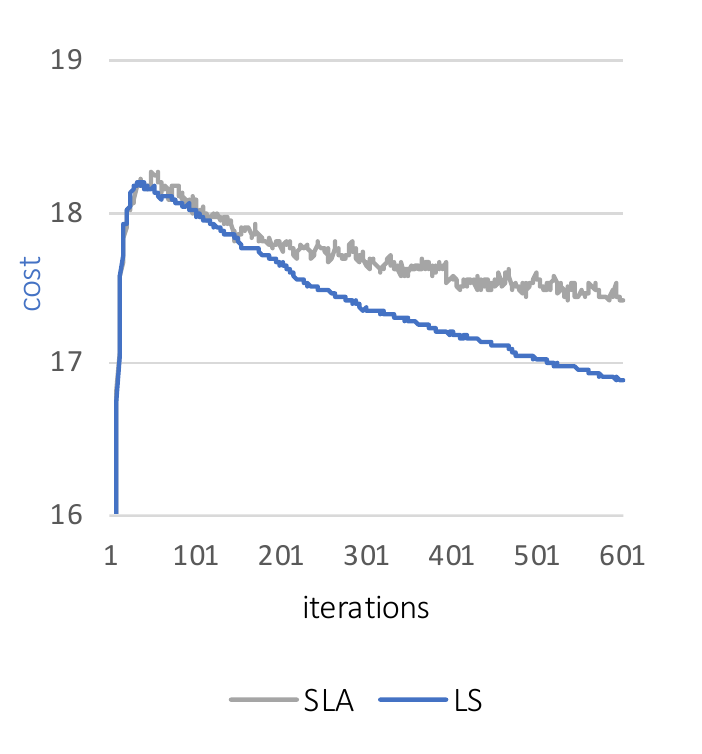}
			\caption{cost curve for SLA/LS.}
		\end{subfigure}
		\begin{subfigure}{0.32\textwidth}
			\includegraphics[width=\linewidth]{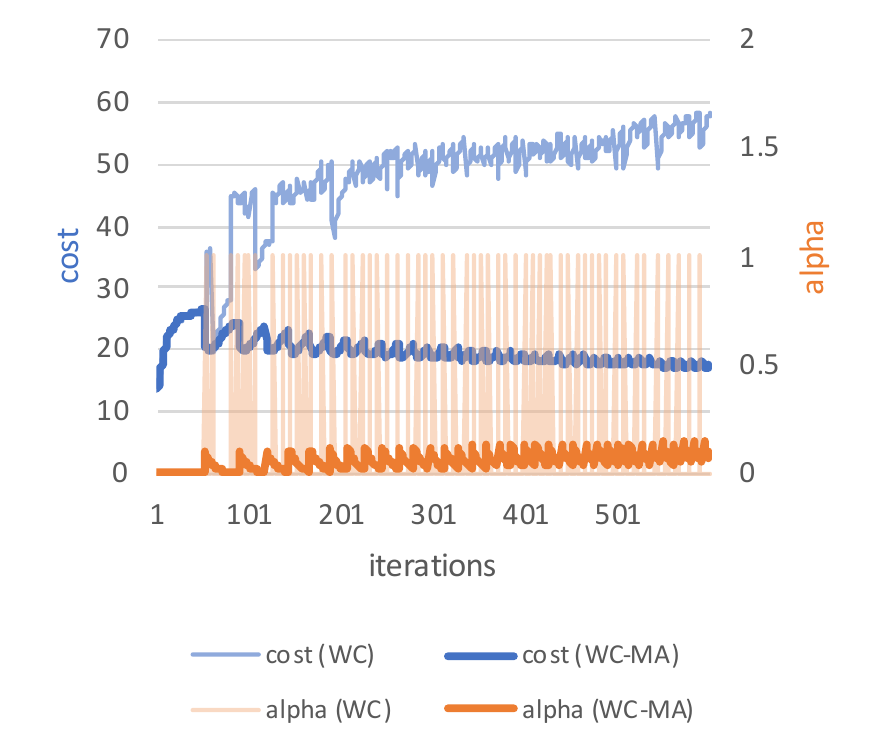}
			\caption{cost curve for WC.}
		\end{subfigure}
		\begin{subfigure}{0.32\textwidth}
			\includegraphics[width=\linewidth]{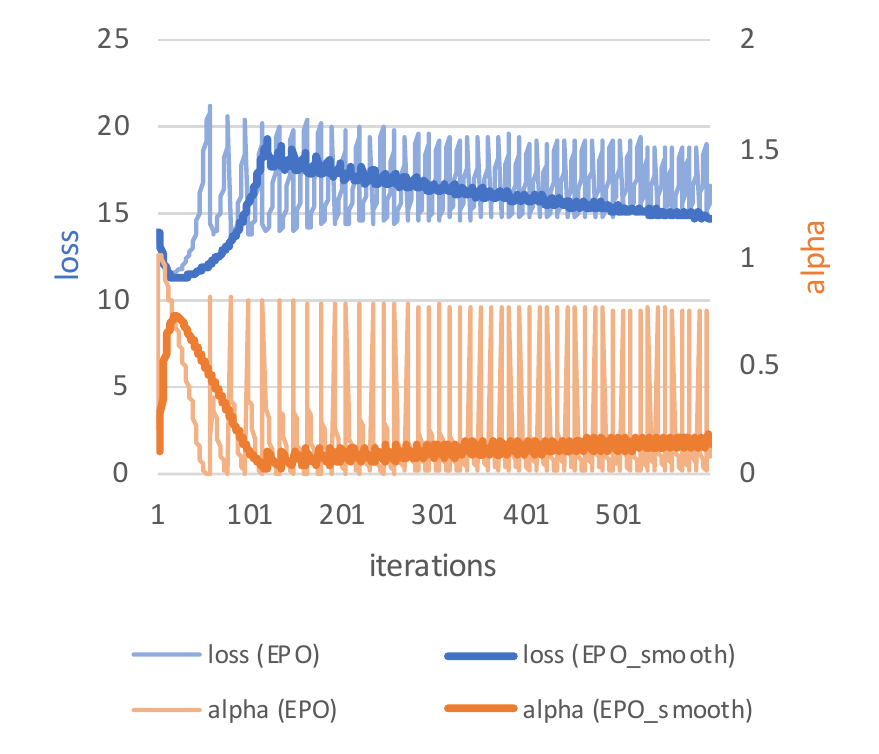}
			\caption{cost curve for EPO.}
		\end{subfigure}
		\caption{Cost curves for (a) SLA/LS, (b) WC and (c) EPO
			for (Click, Rel). For WC and EPO, we also show
			$\alpha$ for Rel. We use light color for the original
			methods and dark for smoothed versions.}
		\label{fig:cost_curves}
	\end{figure*}

\begin{figure*}[htb]
		\centering
		\begin{subfigure}{0.49\textwidth}
			\includegraphics[page=1,
			width=\linewidth]{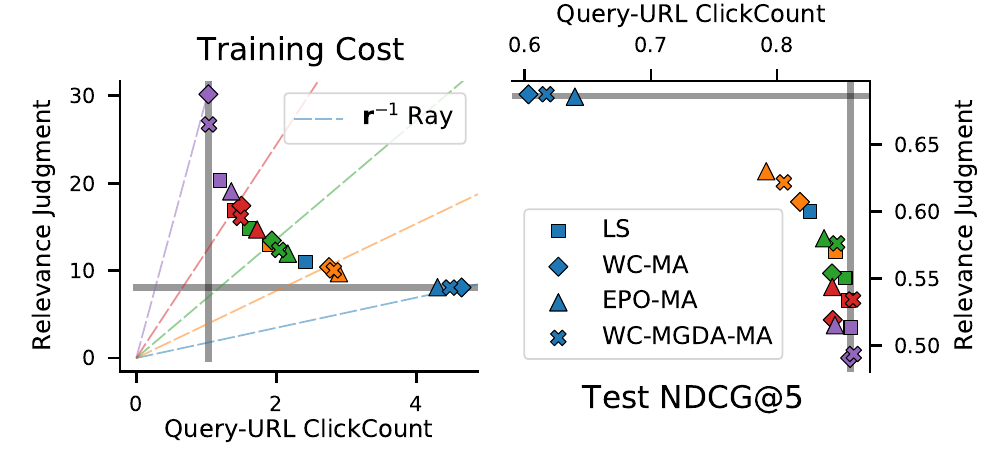}
			\caption{\label{fig:mslr-ma-PD} preference based methods with momving average. }
		\end{subfigure}
		\begin{subfigure}{0.49\textwidth}
			\includegraphics[page=1,
			width=\linewidth]{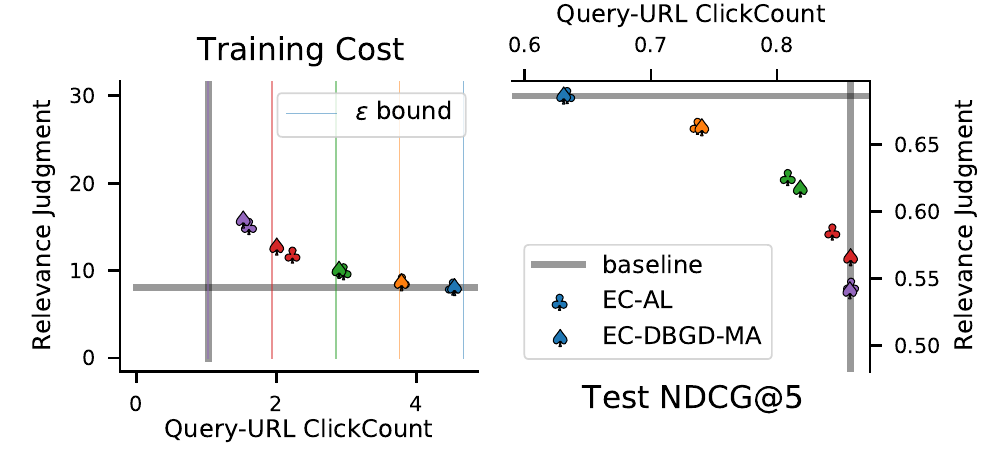}
			\caption{\label{fig:mslr-ma-EC} $\epsilon$-Constraint methods with moving average }
		\end{subfigure}
		\caption{\label{fig:mslr-improved}
			Improved results with moving average. Most of the models are close to the Pareto Front.
		}
	\end{figure*}

	To mitigate the issues,
	we propose to use the moving average (MA) to force smoothing $\boldsymbol{\alpha}$:
	\begin{equation}\label{eq:alpha-smooth}
	\boldsymbol{\alpha}^{t+1} = \nu \boldsymbol{\alpha}^t + (1 - \nu) \boldsymbol{\alpha}^{t-1}, ~ 0 < \nu < 1,
	\end{equation}
	for each $t$, with $ \nu = 0.1 $ througout the paper.
	Cost curves of smoothed versions are shown in Figure \ref{fig:cost_curves} as dark colors for WC and EPO.
	The cost/NDCG result with smoothed $\boldsymbol{\alpha}$ is
	shown in Figure \ref{fig:mslr-improved}. The improvement is
	evident when comparing original methods (smaller mark) and
	smoothed versions (larger mark). After smoothing, the
	models follow similar PF curves as that of LS.                
	
	Hereafter, we quantify the improvement by MA by analyzing
	results over all preferences and randomizations. 
	To compare preference based methods, we use MWL.
	For any methods, we use hypervolume indicator
	(HVI) to quantify PO. For these
	metrics, we use paired t-tests with significance level = 0.05
	to compare methods \footnote{Note, when computing HVI on cost, we scale each cost by the
		worst performance of single objective methods, so the HVI is
		not influenced by different scales of costs. }.
	%
	
	\begin{table*}[ht]
		\caption{\label{tab:ms-biobj}
			Metrics on MSLR and e-commerce 
			dataset for bi-objective experiments.
			``orig'' refers to SLA and original
			versions of WC / EPO / WC-MGDA / EC-DBGD,
			including EC-AL. ``ma'' refers to  LS and
			moving average version of them. Bold numbers mean statistical
			significance between orig and ma. Red number
			refers to a single winner (significance vs. all
			others) for each type.
		}
		\subfloat[MSLR dataset (2-obj)]{                
			\resizebox{0.5\linewidth}{!}{          
				\begin{tabular}{|lrrrrrrrrr|}
					\hline
					\multicolumn{1}{|l|}{\multirow{2}{*}{}}
					& \multicolumn{3}{c|}{MWL (test)}
					& \multicolumn{3}{c|}{HVI (train cost)}
					& \multicolumn{3}{c|}{HVI (test NDCG)}                                                            \\ \cline{2-10} 
					\multicolumn{1}{|l|}{}                  & \multicolumn{1}{c|}{orig} & \multicolumn{1}{c|}{ma} & \multicolumn{1}{c|}{gain\%} & \multicolumn{1}{c|}{orig} & \multicolumn{1}{c|}{ma} & \multicolumn{1}{c|}{gain\%} & \multicolumn{1}{c|}{orig} & \multicolumn{1}{c|}{ma} & \multicolumn{1}{c|}{gain\%} \\ \hline
					\multicolumn{10}{|c|}{Preference based}                                                                                                                                                                                                                                                                              \\ \hline
					\multicolumn{1}{|c|}{SLA/LS} & \multicolumn{1}{r|}{2.24}           & \multicolumn{1}{r|}{\textbf{2.09}}       &  \multicolumn{1}{r|}{-6.7\%}       & \multicolumn{1}{r|}{3.51}           & \multicolumn{1}{r|}{\textbf{3.55}}       &    \multicolumn{1}{r|} {1.0\%}      & \multicolumn{1}{r|}{0.93}           & \multicolumn{1}{r|}{\textbf{0.96}}       &       \multicolumn{1}{r|}{2.2\%}     \\ \hline
					\multicolumn{1}{|c|}{WC}  & \multicolumn{1}{r|}{5.08}           & \multicolumn{1}{r|}{\textbf{1.97}}       &    \multicolumn{1}{r|}{-61.7\%}   & \multicolumn{1}{r|}{3.40}           & \multicolumn{1}{r|}{\textbf{3.55}}       &    \multicolumn{1}{r|}{4.5\%}     & \multicolumn{1}{r|}{0.95}           & \multicolumn{1}{r|}{\textbf{0.96}}       &         \multicolumn{1}{r|}{2.1\%}         \\ \hline
					\multicolumn{1}{|c|}{EPO}   & \multicolumn{1}{r|}{2.55}           & \multicolumn{1}{r|}{\textbf{2.02}}       &    \multicolumn{1}{r|} {-20.7\%}    & \multicolumn{1}{r|}{3.51}           & \multicolumn{1}{r|}{\textbf{3.56}}       &     \multicolumn{1}{r|} {1.4\%}     & \multicolumn{1}{r|}{0.95}           & \multicolumn{1}{r|}{\textbf{0.97}}       &          \multicolumn{1}{r|}{1.6\%}         \\ \hline  
					\multicolumn{1}{|c|}{WC-MGDA}   & \multicolumn{1}{r|}{2.02}           & \multicolumn{1}{r|}{\textcolor{red}{\textbf{1.93}}}       &    \multicolumn{1}{r|} {-4.6\%}    & \multicolumn{1}{r|}{3.53}           & \multicolumn{1}{r|}{\textbf{3.57}}       &     \multicolumn{1}{r|} {1.0\%}     & \multicolumn{1}{r|}{0.96}           & \multicolumn{1}{r|}{\textbf{0.97}}       &          \multicolumn{1}{r|}{0.9\%}         \\ \hline  
					\multicolumn{10}{|c|}{EC method}                                                                                                                                                                                                                                                                              \\ \hline
					\multicolumn{1}{|c|}{EC-AL}  &
					\multicolumn{1}{r|}{--}   & \multicolumn{1}{r|}{--}       &  \multicolumn{1}{r|}{--}
					& \multicolumn{1}{r|}{3.52}           & \multicolumn{1}{r|}{--}       &    \multicolumn{1}{r|} {--}      & \multicolumn{1}{r|}{0.97} & \multicolumn{1}{r|}{--}       &        \multicolumn{1}{r|}{--}     \\ \hline
					\multicolumn{1}{|c|}{EC-DBGD}    &
					\multicolumn{1}{r|}{--}   & \multicolumn{1}{r|}{--}       &    \multicolumn{1}{r|}{--}   & \multicolumn{1}{r|}{3.47}     & \multicolumn{1}{r|}{\textbf{3.52}}   &  \multicolumn{1}{r|}{1.5\%}     & \multicolumn{1}{r|}{0.95} & \multicolumn{1}{r|}{\textbf{0.97}} & \multicolumn{1}{r|}{1.8\%}         \\ \hline
					

				\end{tabular}
			}
		}
		~
		\subfloat[e-commerce  dataset (2-obj)]{
			\resizebox{0.5\linewidth}{!}{          
				\begin{tabular}{|lrrrrrrrrr|}
					\hline
					\multicolumn{1}{|l|}{\multirow{2}{*}{}}
					& \multicolumn{3}{c|}{MWL (test)}
					& \multicolumn{3}{c|}{HVI (train cost)}
					& \multicolumn{3}{c|}{HVI (test NDCG)}                                                            \\ \cline{2-10} 
					\multicolumn{1}{|l|}{}                  & \multicolumn{1}{c|}{orig} & \multicolumn{1}{c|}{ma} & \multicolumn{1}{c|}{gain\%} & \multicolumn{1}{c|}{orig} & \multicolumn{1}{c|}{ma} & \multicolumn{1}{c|}{gain\%} & \multicolumn{1}{c|}{orig} & \multicolumn{1}{c|}{ma} & \multicolumn{1}{c|}{gain\%} \\ \hline
					\multicolumn{10}{|c|}{Preference
						based}                                                                                                                                                                                                                                                                              \\ \hline
					\multicolumn{1}{|c|}{SLA/LS}    & \multicolumn{1}{r|}{2.14}    & \multicolumn{1}{r|}{\textbf{2.14}}      & \multicolumn{1}{r|}{-1.7\%} & \multicolumn{1}{r|}{2.89}    & \multicolumn{1}{r|}{\textbf{2.90}}      & \multicolumn{1}{r|}{0.6\%}     & \multicolumn{1}{r|}{0.94}    & \multicolumn{1}{r|}{\textbf{0.95}}      & 1.27\%                            \\ \hline  
					\multicolumn{1}{|c|}{WC}   & \multicolumn{1}{r|}{15.7}     & \multicolumn{1}{r|}{{\textbf{2.05}}}       &   \multicolumn{1}{r|}{-86.9\%}  & \multicolumn{1}{r|}{2.65}     & \multicolumn{1}{r|}{{\textbf{2.94}} }      &  \multicolumn{1}{r|}{ 11.0\%}   & \multicolumn{1}{r|}{0.93}     & \multicolumn{1}{r|}{{\textbf{0.98}}}       &   5.39\%                           \\ \hline
					\multicolumn{1}{|c|}{EPO}   & \multicolumn{1}{r|}{6.15}     & \multicolumn{1}{r|}{\textbf{2.22}}       &  \multicolumn{1}{r|}{-63.9\%}   & \multicolumn{1}{r|}{2.85}     & \multicolumn{1}{r|}{{\textbf{2.90}}}       &   \multicolumn{1}{r|}{ 1.7\%}                 & \multicolumn{1}{r|}{0.94}     & \multicolumn{1}{r|}{\textbf{0.96}}       &      2.18\%                       \\ \hline
					\multicolumn{1}{|c|}{WC-MGDA}   & \multicolumn{1}{r|}{5.93}     & \multicolumn{1}{r|}{\textcolor{red}{\textbf{2.03}}}       &  \multicolumn{1}{r|}{-65.9\%}   & \multicolumn{1}{r|}{2.86}     & \multicolumn{1}{r|}{{\textcolor{red}{\textbf{2.96}}}}       &   \multicolumn{1}{r|}{ 3.3\%}                 & \multicolumn{1}{r|}{0.97}     & \multicolumn{1}{r|}{\textcolor{red}{\textbf{0.98}}}       &      1.03\%                       \\ \hline
					\multicolumn{10}{|c|}{EC method}                                                                                                                                                                                                                                                                              \\ \hline
					\multicolumn{1}{|c|}{EC-AL}   & \multicolumn{1}{r|}{--}    & \multicolumn{1}{r|}{--}      & \multicolumn{1}{r|}{--} & \multicolumn{1}{r|}{2.82}    & \multicolumn{1}{r|}{--}      & \multicolumn{1}{r|}{--}     & \multicolumn{1}{r|}{0.97}    & \multicolumn{1}{r|}{--}      & --                            \\ \hline
					\multicolumn{1}{|c|}{EC-DBGD}      & \multicolumn{1}{r|}{--}     & \multicolumn{1}{r|}{--}     &  \multicolumn{1}{r|}{--}  & \multicolumn{1}{r|}{2.88}     & \multicolumn{1}{r|}{\textcolor{red}{\textbf{2.93}}}     &   \multicolumn{1}{r|}{1.5\%}   & \multicolumn{1}{r|}{0.97}     & \multicolumn{1}{r|}{{\textcolor{red}{\textbf{0.98}}}}   &   0.7\%                           \\ \hline
			\end{tabular}}
		}
	\end{table*}
	
	Table \ref{tab:ms-biobj} shows bi-objective results on MSLR and e-commerce 
	datasets.  
	The effect of smoothing is clear. 
	For all cases, the gain due to smoothing is significant for all metrics.
	Notably, it benefits WC significantly -- helping it to become 2nd best
	model behind WC-MGDA. WC-MGDA worked well
	even without MA. When it failed for e-commerce dataset, MA
	helped a lot and make it the best model for all metrics. Overall,
	WC-MGDA showed best performance in MWL and at least competitive performance in
	HVIs.
	For EC methods, EC-DBDA with MA works at
	least as competitive as EC-AL, and significantly better for e-commerce 
	dataset.
	However,  WC-MGDA / EC-DBDA requires extra computation of
	generating gradient matrix while WC / EC-AL does not. 
	Hence, users can choose either methods
	based on the cost-efficiency trade-off.
	For tri-objective experiments, refer to Appendix
	\ref{sec:appendix_mslr_amzn}.
	
	\begin{wrapfigure}{R}{0.5\textwidth}	
		\centering
		\includegraphics[page=1,
		width=1\linewidth]{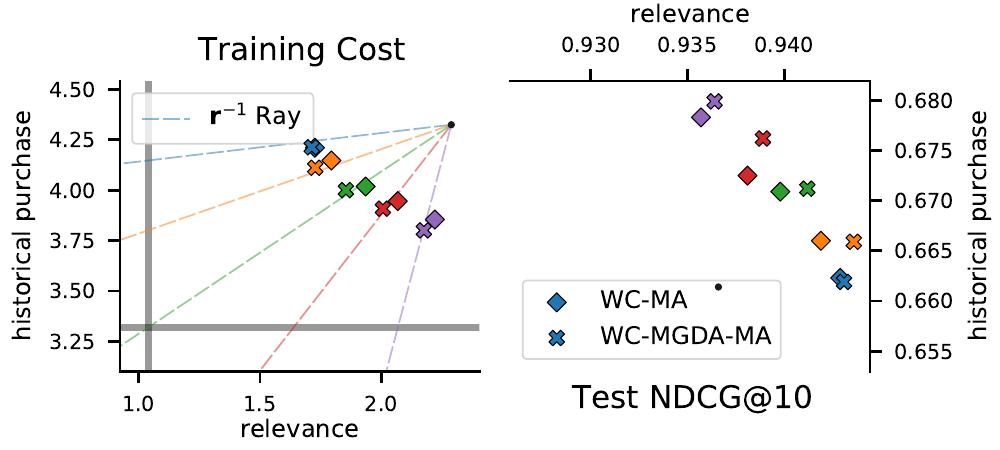}
		\caption{\label{fig:amzn-refpoint} \label{fig:negray}
			Exploring PF from a reference model (black dot) on e-commerce dataset. 
		}
	\end{wrapfigure}

	\subsection{Exploring PF around a reference model on e-commerce dataset}
	We illustrate an important use case of exploring PF around a
	given reference (i.e., pretrained / baseline) model. To simulate a
	reference model, we used LS model with early stopping at 50. 
	We generate equidistributed preference.
	We applied WC and WC-MGDA 
	on this
	setting. Note it is straightforward to modify WC to handle reference
	by subtracting the cost of the reference model. 
	Figure \ref{fig:negray} shows comparison between WC and WC-MGDA. While two methods explore PF from the reference,
	WC-MGDA seems to perform better, which is verified in Appendix \ref{sec:appendix_refp}. Notably, this usage enables us to automatically update production models by a fresh dataset, which improves upon the production model over all objectives.

	\section{Conclusion and future work}
	We develop a general framework for Multi-Label Learning to Rank that is able to consume any first order gradient based MOO algorithm to train a ranking model. We  formalize three types of trade-off specifications, and provide a principled approach to maintain the relative quality of ranking w.r.t. different relevance criteria. We showed SOTA methods perform inferior to simpler baselines, and proposed a remedy, which helped them to achieve Pareto Optimality. We validated our framework using two public datasets and one e-commerce dataset. Furthermore, we showed our framework can also be applied to model auto refresh that improves over all objectives by leveraging preference based method with reference point.
	In future, we plan to explore multiple directions to improve the current MLLTR framework and associated package, such as extending the currently adopted pairwise cost to list-wise, trying for non-convex surrogates that approximate NDCG metric even better than list-wise costs \cite{10.1145/3442381.3449794}, 
	conducting experiments on industry scale datasets, 
	and incorporating more MOO algorithms into the package to further elaborate the sophistication of our framework.
	

	\newpage
	\appendix
	\section{More experiments}\label{appendix:sec:experiments}
	\subsection{3 objectives experiments on MSLR and e-commerce datasets}\label{sec:appendix_mslr_amzn}
	\begin{table}[h]     
		\caption{Evaluation metrics on MSLR dataset: 3 objectives }
		\label{appendix:tab:ms-biobj} 
		\centering
		\resizebox{0.7\linewidth}{!}{          
			\begin{tabular}{|lrrrrrrrrr|}
				\hline
				\multicolumn{1}{|l|}{\multirow{2}{*}{}}
				& \multicolumn{3}{c|}{MWL (test)}
				& \multicolumn{3}{c|}{HI (train cost)}
				& \multicolumn{3}{c|}{HI (test NDCG)}                                                            \\ \cline{2-10} 
				\multicolumn{1}{|l|}{}                  & \multicolumn{1}{c|}{orig} & \multicolumn{1}{c|}{ma} & \multicolumn{1}{c|}{gain\%} & \multicolumn{1}{c|}{orig} & \multicolumn{1}{c|}{ma} & \multicolumn{1}{c|}{gain\%} & \multicolumn{1}{c|}{orig} & \multicolumn{1}{c|}{ma} & \multicolumn{1}{c|}{gain\%} \\ \hline
				\multicolumn{10}{|c|}{Preference based}                                                                                                                                                                                                                                                                              \\ \hline
				\multicolumn{1}{|c|}{SLA/LS}  & \multicolumn{1}{r|}{2.01}   & \multicolumn{1}{r|}{\textbf{1.86}}       &  \multicolumn{1}{r|}{-7.5\%}       & \multicolumn{1}{r|}{6.37}           & \multicolumn{1}{r|}{\textbf{6.52}}       &    \multicolumn{1}{r|} {2.4\%}      & \multicolumn{1}{r|}{0.79} & \multicolumn{1}{r|}{\textbf{0.84}}       &        \multicolumn{1}{r|}{7.0\%}     \\ \hline
				\multicolumn{1}{|c|}{WC}    & \multicolumn{1}{r|}{10.3}   & \multicolumn{1}{r|}{\textbf{1.75}}       &    \multicolumn{1}{r|}{-83.0\%}   & \multicolumn{1}{r|}{6.00}     & \multicolumn{1}{r|}{\textbf{6.57}}   &  \multicolumn{1}{r|}{9.4\%}     & \multicolumn{1}{r|}{0.81} & \multicolumn{1}{r|}{\textbf{0.89}} & \multicolumn{1}{r|}{9.4\%}         \\ \hline
				\multicolumn{1}{|c|}{EPO}   & \multicolumn{1}{r|}{2.46}   & \multicolumn{1}{r|}{\textbf{1.90}} & \multicolumn{1}{r|} {-23.0\%}    & \multicolumn{1}{r|}{6.45}   & \multicolumn{1}{r|}{{\textbf{6.61}}}       &     \multicolumn{1}{r|} {2.4\%}     & \multicolumn{1}{r|}{0.87}    & \multicolumn{1}{r|}{\textbf{0.88}}       &         \multicolumn{1}{r|}{1.7\%}         \\ \hline
				\multicolumn{1}{|c|}{WC-MGDA}   & \multicolumn{1}{r|}{1.88}   & \multicolumn{1}{r|}{\textcolor{red}{\textbf{1.74}}} & \multicolumn{1}{r|} {-7.6\%}    & \multicolumn{1}{r|}{6.54}   & \multicolumn{1}{r|}{{\textbf{6.63}}}       &     \multicolumn{1}{r|} {1.4\%}     & \multicolumn{1}{r|}{0.88}    & \multicolumn{1}{r|}{\textcolor{red}{\textbf{0.90}}}       &         \multicolumn{1}{r|}{2.0\%}         \\ \hline
				\multicolumn{10}{|c|}{EC method}  \\
				\hline
				\multicolumn{1}{|c|}{EC-AL}  &
				\multicolumn{1}{r|}{--}   & \multicolumn{1}{r|}{--}       &  \multicolumn{1}{r|}{--}
				& \multicolumn{1}{r|}{6.51}           & \multicolumn{1}{r|}{--}       &    \multicolumn{1}{r|} {--}      & \multicolumn{1}{r|}{0.88} & \multicolumn{1}{r|}{--}       &        \multicolumn{1}{r|}{--}     \\ \hline
				\multicolumn{1}{|c|}{EC-DBGD}    &
				\multicolumn{1}{r|}{--}   & \multicolumn{1}{r|}{--}       &    \multicolumn{1}{r|}{--}   & \multicolumn{1}{r|}{6.41}     & \multicolumn{1}{r|}{\textbf{6.50}}   &  \multicolumn{1}{r|}{1.4\%}     & \multicolumn{1}{r|}{0.84} & \multicolumn{1}{r|}{\textbf{0.87}} & \multicolumn{1}{r|}{4.1\%}         \\ \hline
				
		\end{tabular}}
	\end{table}

	\begin{table*}[htb]
		\caption{Evaluation metrics on e-commerce dataset: 3 objectives}
		\label{appendix:tab:e-commerce}
		\centering
		
		\resizebox{0.7\linewidth}{!}{          
			\begin{tabular}{|lrrrrrrrrr|}
				\hline
				\multicolumn{1}{|l|}{\multirow{2}{*}{}}
				& \multicolumn{3}{c|}{MWL (test)}
				& \multicolumn{3}{c|}{HI (train cost)}
				& \multicolumn{3}{c|}{HI (test NDCG)}                                                            \\ \cline{2-10} 
				\multicolumn{1}{|l|}{}                  & \multicolumn{1}{c|}{orig} & \multicolumn{1}{c|}{ma} & \multicolumn{1}{c|}{gain\%} & \multicolumn{1}{c|}{orig} & \multicolumn{1}{c|}{ma} & \multicolumn{1}{c|}{gain\%} & \multicolumn{1}{c|}{orig} & \multicolumn{1}{c|}{ma} & \multicolumn{1}{c|}{gain\%} \\ \hline
				\multicolumn{10}{|c|}{Preference based}                                                                                                                                                                                                                                                                              \\ \hline
				\multicolumn{1}{|c|}{SLA/LS}   & \multicolumn{1}{r|}{1.45}    & \multicolumn{1}{r|}{\textbf{1.42}}      & \multicolumn{1}{r|}{-2.2\%} & \multicolumn{1}{r|}{5.14}    & \multicolumn{1}{r|}{\textbf{5.21}}      & \multicolumn{1}{r|}{1.4\%}     & \multicolumn{1}{r|}{0.87}    & \multicolumn{1}{r|}{\textbf{0.90}}      & 1.41\%                            \\ \hline
				\multicolumn{1}{|c|}{WC}      & \multicolumn{1}{r|}{18.4}     & \multicolumn{1}{r|}{{\textbf{1.35}}}     &  \multicolumn{1}{r|}{-92.7\%}  & \multicolumn{1}{r|}{4.07}     & \multicolumn{1}{r|}{\textbf{5.27}}     &   \multicolumn{1}{r|}{29.4\%}   & \multicolumn{1}{r|}{0.86}     & \multicolumn{1}{r|}{{\textbf{0.93}}}   &   8.41\%                           \\ \hline
				\multicolumn{1}{|c|}{EPO}     & \multicolumn{1}{r|}{1.91}     & \multicolumn{1}{r|}{\textbf{1.47}}     & \multicolumn{1}{r|}{-23.1\%}   & \multicolumn{1}{r|}{5.10}     & \multicolumn{1}{r|}{\textbf{5.16}}       &  \multicolumn{1}{r|}{1.2\%}              & \multicolumn{1}{r|}{0.91}     & \multicolumn{1}{r|}{\textbf{0.93}}       &       1.73\%                       \\ \hline
				\multicolumn{1}{|c|}{WC-MGDA}     &
				\multicolumn{1}{r|}{5.83}     & \multicolumn{1}{r|}{\textcolor{red}{\textbf{1.34}}}     & \multicolumn{1}{r|}{-76.9\%}   & \multicolumn{1}{r|}{4.99}     & \multicolumn{1}{r|}{\textcolor{red}{\textbf{5.31}}}       &  \multicolumn{1}{r|}{6.3\%}              & \multicolumn{1}{r|}{0.92}     & \multicolumn{1}{r|}{\textcolor{red}{\textbf{0.94}}}       &       2.08\%                       \\ \hline
				
				\multicolumn{10}{|c|}{EC method}  \\
				\hline
				\multicolumn{1}{|c|}{EC-AL}  &
				\multicolumn{1}{r|}{--}   & \multicolumn{1}{r|}{--}       &  \multicolumn{1}{r|}{--}
				& \multicolumn{1}{r|}{5.00}           & \multicolumn{1}{r|}{--}       &    \multicolumn{1}{r|} {--}      & \multicolumn{1}{r|}{0.93} & \multicolumn{1}{r|}{--}       &        \multicolumn{1}{r|}{--}     \\ \hline
				\multicolumn{1}{|c|}{EC-DBGD}    &
				\multicolumn{1}{r|}{--}   & \multicolumn{1}{r|}{--}       &    \multicolumn{1}{r|}{--}   & \multicolumn{1}{r|}{523}     & \multicolumn{1}{r|}{\textcolor{red}{\textbf{5.35}}}   &  \multicolumn{1}{r|}{2.2\%}     & \multicolumn{1}{r|}{0.93} & \multicolumn{1}{r|}{\textcolor{red}{\textbf{0.95}}} & \multicolumn{1}{r|}{2.7\%}         \\ \hline
				
		\end{tabular}}
	\end{table*}
	
	\subsection{Yahoo Dataset}\label{appendix:tab:yahoo}
	\subsubsection{Yahoo Learning to Rank dataset and experiment settings}
	We experiment on the Yahoo Learning to Rank (YLTR) \cite{pmlr-v14-chapelle11a} challenge dataset with 36K queries.
	Each query-url pair is represented by 700 features. Although
	these features are engineered (not learnt), their
	descriptions, however, are not publicly released.
	Therefore, we selected several labels to use as additional
	objectives.
	Specifically, we selected features that have more than 5
	levels of values, then chose the ones that were least
	correlated among each other.
	In total, we selected 6 objectives including the original
	relevance label, and created 15 bi-objectives and 10
	tri-objective cases. 
	Note as we saw cost vanishing behavior coming from NDCG
	computation within LambdaRank due to low granularity, we use
	RankNet cost \cite{burges2005learning}, which is the pairwise
	cost without NDCG factors.
	For tuning the model hyperparameters, we followed a similar
	strategy as in MSLR-WEB30K, and selected 600 trees and 0.25
	learning rate. We used the original training and test data for our experiment.

	\subsubsection{Remedy by moving average (MA)}
	Tables \ref{appendix:tab:yahoo-2-obj} and \ref{appendix:tab:yahoo-3-obj} show results on
	Yahoo dataset for preference based methods for 2 and 3 objectives, respectively.  Here, bold number means statistically significant gain due to MA. Red indicates the single best model among others. 
	The effect of smoothing is clear. For all cases (except for EPO on
	Yahoo), the gain due to smoothing is significant for all metrics.
	Notably, it benefits WC significantly -- helping it to become 2nd best
	model behind WC-MGDA. For WC-MGDA, it worked well for  Yahoo
	even without MA. Overall,
	WC-MGDA showed best performance in MWL (most important metric in
	preference based methods), and at least competitive performance in
	HVIs. One disadvantage with WC-MGDA is the computational cost to compute
	gradient matrix. To avoid it, WC would be
	a good trade-off. 
	
	\begin{table}[h]     \caption{Evaluation metrics on Yahoo
			dataset: 2 objectives }
		\label{appendix:tab:yahoo-2-obj}
		\centering
		\resizebox{0.7\linewidth}{!}{          
			\begin{tabular}{|lrrrrrrrrr|}
				\hline
				\multicolumn{1}{|l|}{\multirow{2}{*}{}}
				& \multicolumn{3}{c|}{MWL (test)}
				& \multicolumn{3}{c|}{HI (train cost)}
				& \multicolumn{3}{c|}{HI (test NDCG)}                                                            \\ \cline{2-10} 
				\multicolumn{1}{|l|}{}                  & \multicolumn{1}{c|}{orig} & \multicolumn{1}{c|}{ma} & \multicolumn{1}{c|}{gain\%} & \multicolumn{1}{c|}{orig} & \multicolumn{1}{c|}{ma} & \multicolumn{1}{c|}{gain\%} & \multicolumn{1}{c|}{orig} & \multicolumn{1}{c|}{ma} & \multicolumn{1}{c|}{gain\%} \\ \hline
				\multicolumn{10}{|c|}{Preference based}                                                                                                                                                                                                                                                                              \\ \hline
				\multicolumn{1}{|c|}{SLA/LS}               & \multicolumn{1}{r|}{91.7}    & \multicolumn{1}{r|}{\textbf{86.2}}    & \multicolumn{1}{r|}{-6.0\%}      & \multicolumn{1}{r|}{3.28}    & \multicolumn{1}{r|}{\textbf{3.32}}    & \multicolumn{1}{r|}{1.2\%}      & \multicolumn{1}{r|}{0.94}    & \multicolumn{1}{r|}{\textbf{0.95}}    & 1.9\%                       \\ \hline
				\multicolumn{1}{|c|}{WC}                & \multicolumn{1}{r|}{87.3}     & \multicolumn{1}{r|}{{\textbf{81.0}}}     & \multicolumn{1}{r|}{-7.3\%}       & \multicolumn{1}{r|}{3.28}     & \multicolumn{1}{r|}{\textbf{3.34}}     & \multicolumn{1}{r|}{1.9\%}       & \multicolumn{1}{r|}{0.95}     & \multicolumn{1}{r|}{\textbf{0.96}}     &          0.9\%                   \\ \hline
				\multicolumn{1}{|c|}{EPO}               & \multicolumn{1}{r|}{107.3}     & \multicolumn{1}{r|}{107.4}     & \multicolumn{1}{r|}{0.1\%}       & \multicolumn{1}{r|}{3.16}     & \multicolumn{1}{r|}{3.16}     & \multicolumn{1}{r|}{0.0\%}       & \multicolumn{1}{r|}{0.87}     & \multicolumn{1}{r|}{0.87}     &               -0.1\%              \\ \hline
				\multicolumn{1}{|c|}{WC-MGDA}               & \multicolumn{1}{r|}{85.4}     & \multicolumn{1}{r|}{\textcolor{red}{\textbf{80.2}}}     & \multicolumn{1}{r|}{-6.1\%}       & \multicolumn{1}{r|}{3.30}     & \multicolumn{1}{r|}{\textcolor{red}{\textbf{3.35}}}     & \multicolumn{1}{r|}{1.4\%}       & \multicolumn{1}{r|}{0.95}     & \multicolumn{1}{r|}{\textcolor{red}{\textbf{0.96}}}     &               0.7\%              \\ \hline
				\multicolumn{10}{|c|}{EC method}                                                                                                                                                                                                                                                                              \\ \hline
				\multicolumn{1}{|c|}{EC-AL}   & \multicolumn{1}{r|}{--}    & \multicolumn{1}{r|}{--}      & \multicolumn{1}{r|}{--} & \multicolumn{1}{r|}{3.30}    & \multicolumn{1}{r|}{--}      & \multicolumn{1}{r|}{--}     & \multicolumn{1}{r|}{0.95}    & \multicolumn{1}{r|}{--}      & --                            \\ \hline
				\multicolumn{1}{|c|}{EC-DBGD}      & \multicolumn{1}{r|}{--}     & \multicolumn{1}{r|}{--}     &  \multicolumn{1}{r|}{--}  & \multicolumn{1}{r|}{3.29}     & \multicolumn{1}{r|}{{\textbf{3.31}}}     &   \multicolumn{1}{r|}{0.7\%}   & \multicolumn{1}{r|}{0.95}     & \multicolumn{1}{r|}{{{\textbf{0.95}}}}   &   0.4\%                           \\ \hline
			\end{tabular}
		}
	\end{table}

	\begin{table}[h]     \caption{Evaluation metrics on Yahoo
			dataset: 3 objectives}
		\label{appendix:tab:yahoo-3-obj}
		\centering
		\resizebox{0.7\linewidth}{!}{          
			\begin{tabular}{|lrrrrrrrrr|}
				\hline
				\multicolumn{1}{|l|}{\multirow{2}{*}{}}
				& \multicolumn{3}{c|}{MWL (test)}
				& \multicolumn{3}{c|}{HI (train cost)}
				& \multicolumn{3}{c|}{HI (test NDCG)}                                                            \\ \cline{2-10} 
				\multicolumn{1}{|l|}{}                  & \multicolumn{1}{c|}{orig} & \multicolumn{1}{c|}{ma} & \multicolumn{1}{c|}{gain\%} & \multicolumn{1}{c|}{orig} & \multicolumn{1}{c|}{ma} & \multicolumn{1}{c|}{gain\%} & \multicolumn{1}{c|}{orig} & \multicolumn{1}{c|}{ma} & \multicolumn{1}{c|}{gain\%} \\ \hline
				\multicolumn{10}{|c|}{Preference based}                                                                                                                                                                                                                                                                              \\ \hline
				\multicolumn{1}{|c|}{SLA/LS}               & \multicolumn{1}{r|}{70.6}    & \multicolumn{1}{r|}{\textbf{66.7}}    & \multicolumn{1}{r|}{-5.5\%}      & \multicolumn{1}{r|}{6.14}    & \multicolumn{1}{r|}{\textbf{6.23}}    & \multicolumn{1}{r|}{1.4\%}      & \multicolumn{1}{r|}{0.84}    & \multicolumn{1}{r|}{\textbf{0.88}}    & 5.2\%                           \\ \hline
				\multicolumn{1}{|c|}{WC}                & \multicolumn{1}{r|}{72.9}     & \multicolumn{1}{r|}{{\textbf{60.5}}}     & \multicolumn{1}{r|}{-17.0\%}       & \multicolumn{1}{r|}{6.24}     & \multicolumn{1}{r|}{{\textbf{6.40}}}     & \multicolumn{1}{r|}{2.5\%}       & \multicolumn{1}{r|}{0.88}     & \multicolumn{1}{r|}{\textbf{0.91}}     &     3.3\%                        \\ \hline
				\multicolumn{1}{|c|}{EPO}               & \multicolumn{1}{r|}{82.1}     & \multicolumn{1}{r|}{\textbf{82.0}}     & \multicolumn{1}{r|}{-0.1\%}       & \multicolumn{1}{r|}{5.91}     & \multicolumn{1}{r|}{5.90}     & \multicolumn{1}{r|}{-0.1\%}       & \multicolumn{1}{r|}{0.73}     & \multicolumn{1}{r|}{0.73}     &       -0.2\%                      \\ \hline
				\multicolumn{1}{|c|}{WC-MGDA}               & \multicolumn{1}{r|}{66.4}     & \multicolumn{1}{r|}{\textcolor{red}{\textbf{60.3}}}     & \multicolumn{1}{r|}{-9.3\%}       & \multicolumn{1}{r|}{6.31}     & \multicolumn{1}{r|}{\textcolor{red}{\textbf{6.41}}}     & \multicolumn{1}{r|}{1.7\%}       & \multicolumn{1}{r|}{0.89}     & \multicolumn{1}{r|}{\textbf{0.91}}     &       1.9\%                  \\     \hline
				\multicolumn{10}{|c|}{EC method}                                                                                                                                                                                                                                                                              \\ \hline
				\multicolumn{1}{|c|}{EC-AL}   & \multicolumn{1}{r|}{--}    & \multicolumn{1}{r|}{--}      & \multicolumn{1}{r|}{--} & \multicolumn{1}{r|}{6.38}    & \multicolumn{1}{r|}{--}      & \multicolumn{1}{r|}{--}     & \multicolumn{1}{r|}{0.89}    & \multicolumn{1}{r|}{--}      & --                            \\ \hline
				\multicolumn{1}{|c|}{EC-DBGD}      & \multicolumn{1}{r|}{--}     & \multicolumn{1}{r|}{--}     &  \multicolumn{1}{r|}{--}  & \multicolumn{1}{r|}{6.34}     & \multicolumn{1}{r|}{{\textbf{6.39}}}     &   \multicolumn{1}{r|}{0.7\%}   & \multicolumn{1}{r|}{0.89}     & \multicolumn{1}{r|}{{\textcolor{red}{\textbf{0.90}}}}   &   0.3\%                           \\ \hline
			\end{tabular}
		}
	\end{table}
	
	\subsection{Model performance on reference point based methods}\label{sec:appendix_refp}
	Table \ref{tab:negray} shows MWL and HVI for reference point based
	methods (i.e., WC-MGDA and WC). As moving average is proved better, we
	only apply moving average on the methods. We apply them to both e-commerce 
	and MSLR datasets. WC-MGDA clearly beats WC, which is consistent with
	the visualization in Figure \ref{fig:amzn-refpoint}.
	\begin{table}[htb]
		\caption{Metrics on preference with reference points. WC-MGDA (w/ MA) shows significantly better than WC (w/MA)
			in both MWL and HVIs.}
		\label{tab:negray}
		\centering
		\resizebox{0.8\linewidth}{!}{
			\begin{tabular}{|c|rrr|rrr|}
				\hline
				dataset    & \multicolumn{3}{c|}{e-commerce}                                                                             & \multicolumn{3}{c|}{MSLR}                                                                                   \\ \hline
				metric     & \multicolumn{1}{c|}{MWL}              & \multicolumn{1}{c|}{HVI(cost)}         & \multicolumn{1}{c|}{HVI(ndcg)} & \multicolumn{1}{c|}{MWL}              & \multicolumn{1}{c|}{HVI(tr)}         & \multicolumn{1}{c|}{HVI(ndcg)} \\ \hline
				WC-MA      & \multicolumn{1}{r|}{-7.1e-2}          & \multicolumn{1}{r|}{8.9e-3}          & 8.0e-4                       & \multicolumn{1}{r|}{-1.6e-1}          & \multicolumn{1}{r|}{5.1e-2}          & 3.9e-3                       \\ \hline
				WC-MGDA-MA & \multicolumn{1}{r|}{\textcolor{red}{\textbf{-8.9e-2}}} & \multicolumn{1}{r|}{\textcolor{red}{\textbf{1.3e-2}}} & \textcolor{red}{\textbf{9.3e-4}}              & \multicolumn{1}{r|}{\textcolor{red}{\textbf{-1.9e-1}}} & \multicolumn{1}{r|}{\textcolor{red}{\textbf{6.2e-2}}} & \textcolor{red}{\textbf{4.4e-3}}              \\ \hline
				gain (\%)  & \multicolumn{1}{r|}{-27}               & \multicolumn{1}{r|}{50}              & 16                           & \multicolumn{1}{r|}{-19}               & \multicolumn{1}{r|}{21}              & 14                           \\ \hline
		\end{tabular}}
	\end{table}
	
	\bibliographystyle{unsrt}
	\bibliography{references}
\end{document}